# An innovative, fast and facile soft-template approach for the fabrication of porous PDMS for oil-water separation

A. Turco[a†], E. Primiceri[b], M. Frigione[c], G. Maruccio[b], C. Malitesta[a]

Oil wastewater and spilled oil caused serious environmental pollution and damage to public health in the last years. Therefore, considerable efforts are made to develop sorbent materials able to separate oil from water with high selectivity and sorption capacity. However most of them are low reusable, with low volume absorption capacity and poor mechanical properties. Moreover, the synthesis is time-consuming, complex and expensive limiting its practical application in case of emergency. Here we propose an innovative approach for the fabrication of porous PDMS starting from an inverse water-in-silicone procedure able to selectively collect oil from water in few seconds. The synthesis is dramatically faster than previous approaches, permitting the fabrication of the material in few minutes independently from the dimension of the sponges. The porous material evidenced a higher volume sorption capacity with respect to other materials already proposed for oil sorption from water and excellent mechanical and reusability properties.This innovative fast and simple approach can be successful in case of emergency, as oil spill accidents, permitting *in situ* fabrication of porous absorbents..

## 1. Introduction

Crude oil has been fundamental for human life from the beginning of industrial civilization. However, in recent years, oil spill accidents and leakage of organic liquids determined a huge damage to ecological systems with consequences in health, social and economic fields[1–3]. An example is the oil leakage in Gulf of Mexico in 2010, when an estimated release of 4.9 million barrels of crude oil into the ocean occured causing a colossal disaster for marine species, with long-term effects still evaluable today[4]. Therefore, a wide variety of approaches has been developed to separate oil from water, such as physical sorption by sorbent materials, mechanical recovery by oil skimmers[5,6], in situ burning[7], physical diffusion[8], filtration membranes[9,10], centrifugation[11] and biodegradation.[12]. However, their practical use still remains problematic due to the low separation efficiency, the use of cumbersome instrumentation and/or the generation of secondary pollutants. To solve these problems, the synthesis of lightweight porous oil absorbent with high volume absorption capacity is one of the most promising strategy[13]. These materials can separate oil from water and concentrate organic liquids inside the pores, permitting their transport and storage. Practical application also need easy, fast and cheap fabrication and the possibility of the sorbent material to be reusable to cut the costs of oil spill recovery[14]. Different materials have been proposed for this aim, such as carbon nanotubes and/or graphene sponges[15–20], polystyrene fibers[21], polyurethane sponges[17,22] and nanocomposite materials[13]. Despite their good performances in water oil separation and mass sorption capacity, most of the materials present drawbacks as poor durability, complex and time-consuming fabrication and poor volume absorption capacity.

Polydimethylsiloxane (PDMS) is a polymeric material used in many research fields due to its excellent characteristics such as high hydrophobicity and flexibility, oleophilicity, thermal, mechanical and chemical stability and easy fabrication. Recently this material has been proposed in porous form as a good candidate for water oil separation. The most widely-adopted method to obtain this material is the sacrificial hard template approach proposed for the first time by Choi et al.[23] in 2011 where the authors fabricated reusable PDMS sponges with fast oil absorption using sugar particles as template. The synthetic steps included shaping of sugar templates and the use of vacuum apparatus that complicate the large-scale production of the material. To solve these problems Zhang et al.[24] proposed an innovative synthesis in which hard template was easily added to PDMS prepolymer diluted in p-xylene without the use of expensive equipment. Despite the good oil absorption properties of the sponges the synthesis still require a long procedure to remove hard template from the polymeric matrix. Zhao et al.[25] synthesized sponges with high oil absorption by diluting PDMS prepolymer in dimethicone and using sugar particles as template. Also in this case, the large-scale production of the material was prevented by the required use of centrifuges and long procedure. To speed up the synthetic procedure more recently Yu et al[26] tried to substitute sugar particles with citric acid monohydate (CAM) as hard template and dissolved them in ethanol. Taking advantage of the superwettability of ethanol to the porous PDMS, CAM could be easily removed from the porous PDMS in shorter time compared to previous approaches. However at least 6 hours for a small piece of the material are still necessary to remove CAM completely. However, due to diffusion of the solvent inside the polymeric pores for the CAM solubilization, longer time is needed for bigger sponges. Recently our group proposed an approach in which PDMS prepolymer was infiltrated between glucose crystals packed in a syringe, applying a positive pressure on their surface. These sponges revealed excellent and improved volume absorption capacity with respect other oil absorbents without the use of cumbersome instrumentations although the overall process still remains time-consuming[13]. Consequently, all the proposed approaches appear time-consuming and require environment-harmful solvents making difficult their real application in case of emergency, where fast material preparation and simple operations are required. Moreover, most of them need the use of cumbersome instrumentations for the preparation such as centrifugation or vacuum operation that further complicate sponges production. An attractive route to prepare highly porous and permeable polymeric materials is represented by emulsion templating method in which a high internal phase emulsion (HIPE) is prepared. The external phase is converted to polymer and the emulsion droplet are removed yielding a highly-interconnected network of micron sized pores. Recently, a few of different attempts have been performed to prepare microporous PDMS by HIPE techniques. In 2015 Tebboth and collaborators[27] produced an elastomeric PDMS containing an aqueous solution of hydrogen carbonate ($NaHCO_3$). Subsequently thermal decomposition of $NaHCO_3$ cause the release of carbon dioxide into the polymer structure. Once placed the polymer under reduced pressure it can expand to many time their original size. Kovalenko et al.[28] prepared soft porous PDMS by the UV polymerization of inverse water-in-silicone PDMS emulsion in presence of surfactants. The use of different reagents, cumbersome instrumentations, long time synthesis limited the large-scale synthesis and consequently the application of both materials for water/oil separation. Here we propose an innovative approach for the facile synthesis of highly porous PDMS materials starting from an inverse water-in-silicone procedure similarly to a modified emulsion templating technique[29]. However we did not observe the presence of voids and windows inside the pores characteristic of emulsion templating approach[30]. The synthesis is completed in few minutes and the obtained sponges presented swelling properties and interconnected pores that favour absorption and retention of various organic liquids with a high and fast absorbency rate. Without the shaping of hard-template, vacuum or centrifugation operation, the preparation strategy is straightforward and dramatically faster than previous approaches. Moreover, the organic liquids can be easily recovered by simply squeezing the sponges that can be thus reused without loss of efficiency for hundreds of cycles.

## Experimental section

### Chemicals

The PDMS prepolymer (Sylgard 184) and a curing agent were purchased from Dow Corning. Hexane, dichloromethane, chloroform, toluene, tetrahydrofuran, petroleum ether, ether, and glucose were purchased from Sigma Aldrich and used as received. Gasoline was purchased from Kuwait Petroleum Corporation.

### Preparation of PDMS-MWNTs sponges

Porous PDMS sponges were prepared according to the following procedure. Unless otherwise indicated a continuous phase consisting of PDMS prepolymer and the thermal curing agent in a ratio of 10:1 by weight were placed into a polypropylene tube and were diluted in hexane with a ratio solvent/prepolymer equal to 8:2 and intensively stirred for 5 minutes. Then 50% wt. of Milli-Q water was added drop by drop under continuous stirring in the continuous phase. After that the solution was further mixed for 20 minutes more. After stirring, the solution had the form of a viscous liquid. The tube containing the as prepared emulsion was then placed in an oven at 120°C for ten minutes to accomplish the polymerization. Therefore, the obtained 3D porous PDMS was removed from the polypropylene tube and used for the experiments. Different PDMS sponges were prepared by changing different parameters such as the concentration of the prepolymer, solvent to dilute the pre-polymer, solvent used as dispersed phase, and the amount of dispersed phase.

### Oil absorbency and reusability of PDMS sponge

A piece of sample was immersed in a bath containing only oil at room temperature for 20 minutes. After that, the sample was removed from oil, wiped with filter paper in order to remove excess oil and weighed. The oil mass sorption capacity ($M_{abs}$) and volume absorption ($V_{abs}$) capacity were evaluated by using the following equation:

$$M_{abs} = (m - m_0)/m_0$$
$$V_{abs} = (m - m_0)\, \rho_0/\rho m_0$$

where m is the weight of the sample after sorption, $m_0$ is the initial weight of the sample, $\rho_0$ and $\rho$ are density values of the absorbent material and absorbed oil, respectively. Repeated absorption–desorption cycles of oils were performed to evaluate sponge reusability. For this purpose, the sample was immersed in oil until the absorption equilibrium was reached (i.e. 20 min), and then weighed to calculate the oil mass sorption capacity. The sample was then squeezed and washed with ethanol three times and dried in an oven at 60 °C. The absorption–desorption procedure was repeated 3 times for each tested oil.

### Characterization of the PDMS sponges

The dynamic and static contact angle measurements were carried out using an OCA 15 Plus instrument (dataPhysiscs Instruments, Filderstadt, Germany) equipped with a high resolution camera and an automated liquid dispenser. SCA 20 software was employed to obtain additional information on the samples; the software uses an algorithm based on the Young Laplace equation and it allows to correlate the shape of the drop with its surface tension and to measure the contact angle between the liquid and the analyzed surface.

Morphological characterization of PDMS sponges samples was carried out by an inverted optical microscope (NIKON mod. DS-5MC camera) in bright field mode. Nikon NIS-Elements ND2 software system has been used to measure pore size. Quasi-static hysteresis compression tests were performed on a LLOYD LR50K Plus dynamometer equipped with 50 mm diameter parallel plate tools and a 50 kN load cell. Tests were performed by loading the samples at different strain levels, equal to 60 and 90%. The loading and unloading stages were performed at a rate of deformation of 5 mm*min$^{-1}$ and 10 mm*min$^{-1}$ for 60% and 90% strain, respectively.

The porosity ($\Phi$) was measured with a methanol saturation method[24] according to the following equation:

$$\Phi = \gamma_{sat} - \gamma_{dry}/\gamma_{meth}$$

Where $\gamma_{sat}$, $\gamma_{dry}$ and $\gamma_{meth}$ indicate the densities of saturated PDMS, dry oil absorbent, and methanol, respectively.

## Results and discussion

### Preparation of PDMS sponge

PDMS elastomer was chosen because of its intrinsic properties such as elasticity, mechanical stability and hydrophobicity. PDMS oil absorbents were synthesized starting from an inverse water-in-silicone procedure (Figure 1a). Compared with the conventional sugar-template methods, this approach avoids the use of vacuum operation or centrifugation, the preparation of shaped salt template beforehand and long washing procedure necessary for hard template removal. An appropriate amount of water acting as soft template was added to dilute prepolymer drop by drop under continuous stirring to produce an emulsion. Then the emulsion was cured in oven at 120°C. The emulsion was stable after the formation during all the synthetic steps. No noticeable phase separation was observed during the time necessary between emulsion formation and curing of the dissolved silicone. At this temperature PDMS prepolymer quickly form cross-linked bonding between PDMS prepolymer chains, giving a semi-rigid gel that surrounds emulsion droplets in few minutes. As a consequence, closed-pores entrapping soft template are formed. During evaporation that naturally occur at that temperature, the soft template increase its volume and physically enlarge the closed-pore formed by not completely polymerized PDMS. When the pressure due to evaporation of soft template inside the formed closed pores becomes high enough, the interconnection between PDMS prepolymer chains could be physically broken causing the explosion of the closed-pores leaving large open cavities inside the polymeric matrices with an average dimension of 406±302 µm, a minimum pore dimension of 65.8 µm and a maximum of 1.768 mm (Figure 1b-c and figure S1).

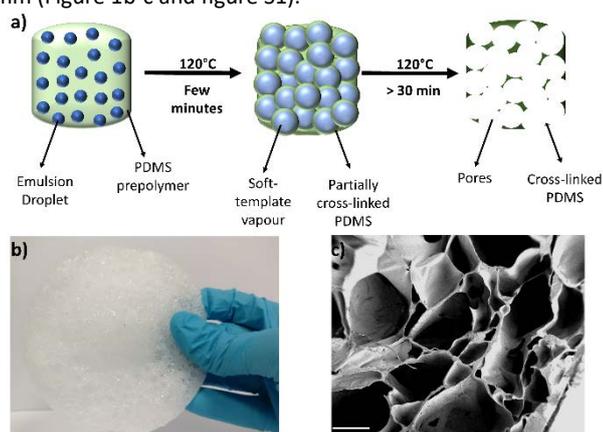

Figure 1 (a) Schematic illustration of the preparation of the PDMS sponge. (b) photograph of the freshly prepared PDMS sponge. (c) SEM image of a sponge portion (scale bar=500µm)

Therefore, all the prepolymer was converted in a porous structure with interconnected pores in few minutes (20 min) independently from the dimension of the sponge (Figure 1b).
At the end of the process both diluting and template solvent are evaporated, leaving porous PDMS structure ready to use. The simple preparation method here adopted, makes sponges fabrication suitable for large-scale production without the need of complex, expensive and cumbersome instrumentation. Moreover, the preparation is faster (few minutes) thanother reported methods used for porous PDMS fabrication for which many hours or days are required to produce large amount of the material.

**Performance of PDMS sponge in oil absorbency**
For an ideal oil absorbent, a material highly porous and with 3D interconnected pores is necessary. It is well known that porous PDMS has the ability to take up oil[13,23–26].
To select the optimal conditions, determining the most satisfactory oil uptake properties, different solvents and amount for both prepolymer solubilization and acting as soft template have been used for the preparation of PDMS sponges.
Dichloromethane (DCM) absorption has been used to evaluate oil uptake properties of different prepared sponges. Figure 2 shows porosity and dichloromethane absorbency of the sponges prepared using water as soft template at 50% in weight with respect to PDMS prepolymer and hexane in different ratio. Once dipped in contact with organic solvent all the prepared sponges increased their volume during the absorption process, indicating the occurrence of a swelling process (Figure 2b). As reported, an oil mass absorption higher than 21 g g$^{-1}$ with a swell ratio of 3.2 v/v$_0$ for dichloromethane was obtained for the PDMS sponges prepared with a ratio of 8:2 m$_{PDMS}$/m$_{Hexane}$. This sponge is characterized by a porosity of about 93%, higher than other PDMS porous material for which a maximum of 84% of porosity was observed[24]. Both higher and lower dilution of prepolymer in hexane lead to a decrease in porosity and consequently in dichloromethane absorbency. This behaviour could be explained by two different mechanisms. It is well known that non-diluted prepolymer has a higher cross-linking degree that avoid swelling of the polymer during oil uptake process decreasing its absorbency[24,25]. If more hexane is required to dilute prepolymer, less is the cross-linking degree of the PDMS, therefore increasing hexane amount more swellable sponges with higher absorption are obtained. Moreover, non-diluted prepolymer could polymerize faster and with a higher cross-linking degree preventing pore enlargement during evaporation of the soft-template addressing smaller porosity. However, it is clear as increasing m$_{PDMS}$/m$_{Hexane}$ for values higher than 8:2 leads to a decrease of the absorption capacity. We hypothesized that if the amount of hexane increases over a certain amount, the cross-linking degree of PDMS sponges decreases dramatically, consequently the skeleton could not support its own weight and collapses[24]. This causes a decrease in the porosity of the sponges and consequently in the absorption capacity. As a further proof, we observed that for m$_{PDMS}$/m$_{Hexane}$ ratio lower than 4:6 the sponge was completely collapsed sticky and not suitable for oil absorption.

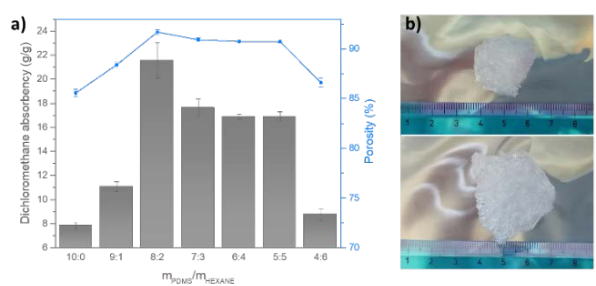

Figure 2 a) Dichloromethane mass absorption (g g-1) (columns) and porosity (dots) of the PDMS sponges with different m$_{PDMS}$/m$_{Hexane}$ ratios. b) Swelling process of PDMS sponge before (top) and after (bottom) immersion in oil.

The impact of the solvent used in PDMS dilution during PDMS sponge formation was evaluated by preparing different sponges with different solvents. We tested hexane, toluene and dichloromethane; the dilution ratio was 2:8 solvent /PDMS prepolymer in presence of 50% of water with respect to prepolymer and the uptake of dichloromethane was evaluated. The solvent used to dilute the prepolymer has a strong effect on the oil uptake process (Figure 3) and lower is the polarity of the solvent, higher is dichloromethane absorbency. Arguably, the higher the polarity of the solvent, the greater is its ability to form hydrogen bonds with water. This could cause a decrease in the porosity of the sponges as already reported for systems based on polymerization starting from an emulsion[31].

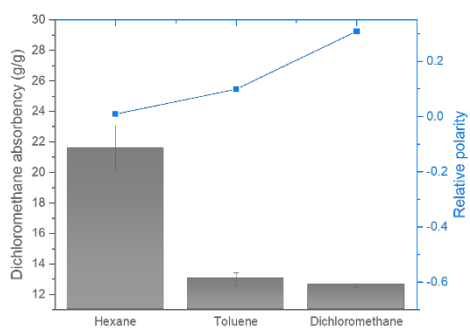

Figure 3 Variation of mass absorption (g g$^{-1}$) of the PDMS sponges for dichloromethane with different solvent used to dilute the PDMS prepolymer.

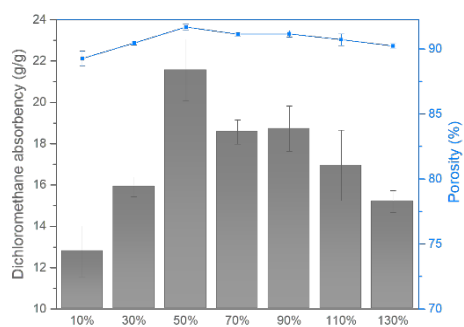

Figure 5 Dichloromethane mass absorption (g g-1) (columns) and porosity (dots) of the PDMS sponges prepared with different amount in wt with respect PDMS prepolymer.

In hard template synthesis of PDMS sponges was already demonstrated as different template could cause different porosity[13]. In the present work, the template namely soft-template is represented by a solvent immiscible with prepolymer diluted in opportune solvent. Sponges with different soft template at 50% in weight with respect PDMS prepolymer have been prepared to evaluate the impact of them on dichloromethane absorption. The $m_{PDMS}/m_{Hexane}$ was kept at a ratio of 8:2. Figure 4 shows as the soft-template has a significant effect on the oil uptake process. We observed that oil absorption increases increasing the boiling point of soft-template. We hypothesized that this could be due since polymerization occurs at high temperature, then part of the soft template could evaporate before the formation of prepolymer cross-linked bonding. Consequently, smaller emulsion droplet will be present in the mixture causing the formation of less porous sponges.

In figure 6 normalized dichloromethane absorption of the sponges with the higher dichloromethane absorbency is plotted as a function of time. The maximum absorbency is observed in less than 30 seconds. The absorption rate is comparable with other PDMS based materials[24] and higher with respect to other gel oil absorbents, which reached the equilibrium in 2 hours or even longer[32–34].

Moreover, different sponges produced in different times exhibited similar dichloromethane uptake suggesting that the new synthetic route produces material with reproducible oil uptake properties (RSD=10% n=3).

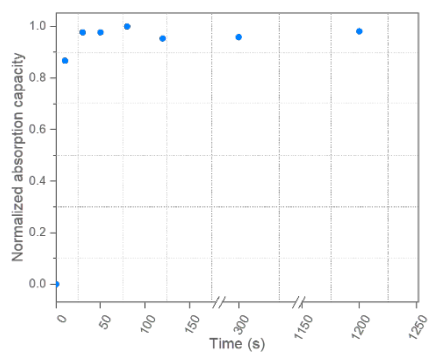

Figure 6 Oil mass absorption for different contact times of PDMS sponge.

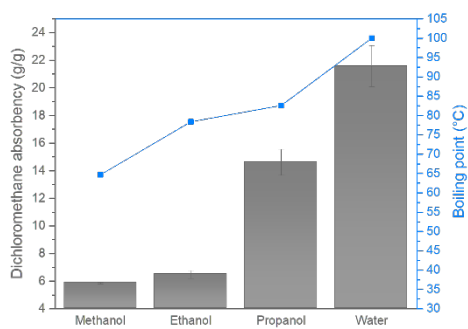

Figure 4 Variation of mass absorption (g g$^{-1}$) of the PDMS sponges prepared from different soft-template.

Similarly, a decrease in oil absorbency is observed if less of 50% soft template with respect PDMS prepolymer was used (figure 5). However, also increasing soft template amount to values higher than 50% wt compared to prepolymer lead to a decrease in dichloromethane absorbency and porosity. We supposed that higher is the amount of template solvent larger is the distance between the forming polymeric chains causing a decrease of the crosslinking degree. Consequently, the skeleton collapses on its weight decreasing the porosity.

Importantly the as-obtained PDMS sponges have intrinsic hydrophobicity and oleophilicity, highlighted by water contact angle (CA) of 145.5±1° (Figure 7a). Thanks to lightweight, porous structure and hydrophobicity properties, PDMS sponge float on water surface and when immersed in water with the help of an external force, an air cushion around the sponge that maintain the sponge dryness is observed (Figure 7b). This is evident in Movie S1 in which paper remained dried after that a sponge previously submerged in water was rest on its surface. Moreover, the weight of the sponge remains constant before and after immersion in water. The hydrophobicity of the PDMS sponge is maintained also towards corrosive aqueous liquid including 1 M HCl and 2 M NaOH with a water CA of ~144.7° and ~146.4°. The combined hydrophobicity and oleophilicity of the sponge are commonly attributed to the low surface free energy of the PDMS material and the porous structure. Normally bulk PDMS showed excellent hydrophobicity with a water contact angle of 105°[35]. It is well known that porous morphology can increase the water contact angle of hydrophobic surface due to air entrapment, as described by the Cassie-Baxter wetting model[36].

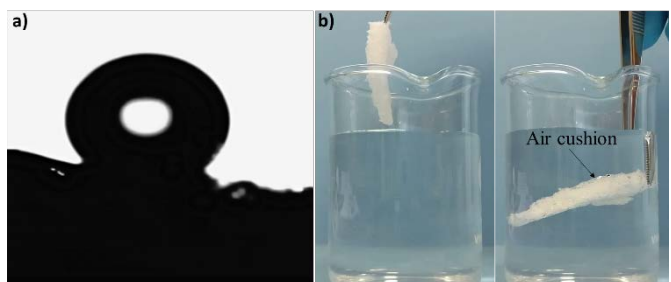

Figure 7 a) Digital image of a drop of water deposited on PDMS. b) PDMS sponge during immersion in water.

Mechanical properties of sorbent materials are important for practical application since in materials with lower mechanical strength the viable stack height of the sorbent after absorption is only a few centimeters.

PDMS is well known for its excellent compression properties in both porous and bulk form. Figure 8a reports the stress-strain curve under a loading-unloading compression cycle for the sponge with the higher dichloromethane uptake. The modulus increases slowly until a strain of about 50% a typical performance of soft foam. Nevertheless, values higher than 300 kPa at 60% strain for the sponge are recorded, much higher with respect to most of other reported oil-uptake systems making the material suitable for oil removal[13]. Recyclability of porous materials for water/oil separation is a key point to dramatically cut the costs of production. In figure 8b a PDMS sponge was pressed until 90% of its original volume at 10 mm/min to mimic the squeezing process necessary to remove entrapped oil in real application and normalized absorption capacity after immersion in dichloromethane was recorded. The results show that absorption capacity did not significantly deteriorate until 200 cycles and more than 90% of its absorption capacity remained after 300 runs. The decrease in absorption capacity of PDMS sponges could be ascribed to the small weight changes as deducible from the small decrease of stress value after successive compression cycles at 90% strain necessary to remove absorbed oil (see figure S2). Importantly the 3D porous materials keep their hydrophobicity (CA=145.1 ±1°) after 300 usage cycles evidencing as dichloromethane is not adsorbed on pore surface.

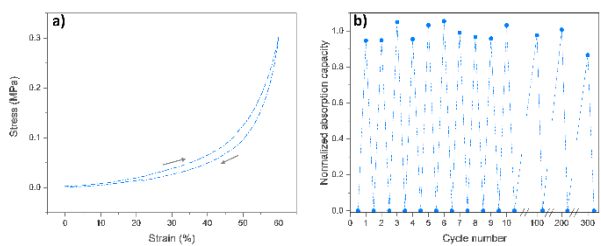

Figure 8 a) Compressive stress-strain curve at 60% strain of PDMS. b) Demonstration of the recyclability of the sponge for dichloromethane absorption.

### Selective absorption of oils from water

With hydrophobic/hydrophilic properties, interconnected porous structure and mechanical stability the PDMS sponges are good candidate as an oil absorbent for water/oil separation. When floating on the surface of an oil water/mixtures, the sponge absorbed in a short time the oil and all water was left (Figure 9a and movie S2) allowing the storage of the collected oil inside the 3D porous structure. Moreover, the PDMS sponge is also able to adsorb heavy oil under water if driven by an external force. It is interesting to note as the weight of the sponges after oil uptake is equal to the weight of the sponge before absorption plus the weight of ~99% of the oil present in water. Moreover, some air bubbles come out from the sponges during submerged oil absorption, confirming the absence of water inside the absorbents after immersion in water. After oil absorption, the sponge float over water surface and can be removed without leakage of oil (Figure 9b and movie S3). This aspect is important in real application. More importantly, the oil stored in the pores can be easily removed by simple squeezing as shown in figure 9c or movie S4.

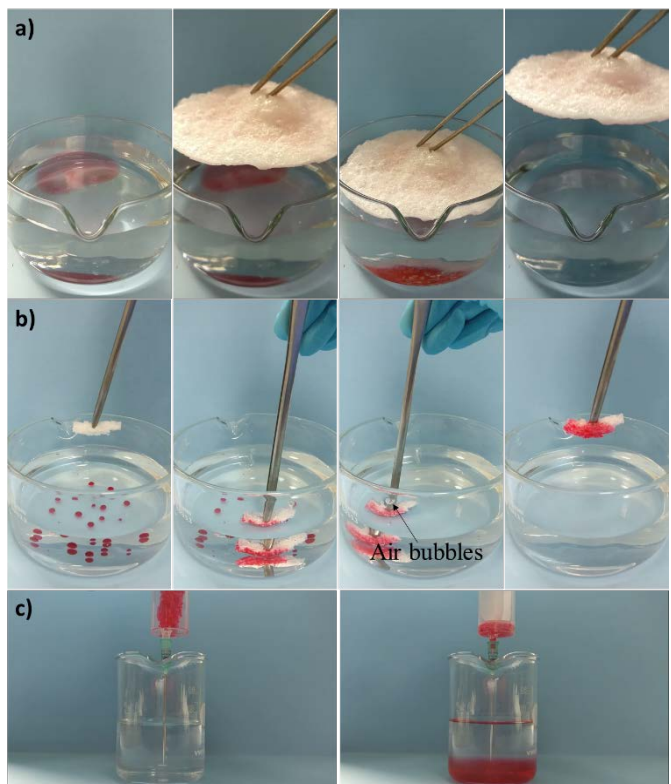

Figure 9 Removal of a) oil red O colored hexane from the water surface and b) oil red O colored chloroform with PDMS sponges. C) recovery of absorbed oil red O colored chloroform from PDMS sponge by squeezing (the transparent solvent is ethanol)

The PDMS oil mass absorption capacity was tested with different oils and organic solvents. As observable in figure 10 the oil mass absorption was in the range of 3.9 to 32.3 g g-1 depending on viscosity, density and surface tension of the organics.

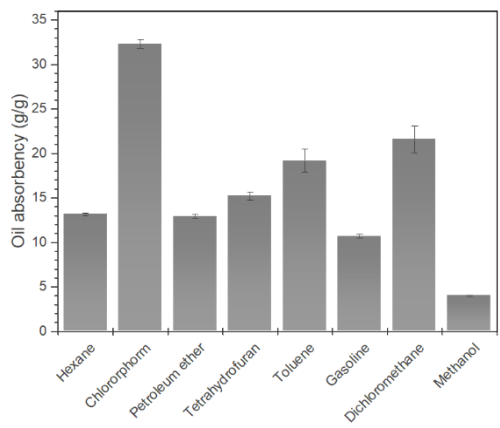

Figure 10 Different oil absorption for PDMS sponge.

It is well known that PDMS based porous materials have a higher volume absorption capacity compared to the other porous oil absorbents such as (polyurethane, carbon soot, aereogel...)[13], this is possible due to the swelling of PDMS after immersion in oil, which significantly increases the available volume for oil storage.

**Table 1** Comparison of various oil absorbents

| Oil absorbent | Water Contact Angle (°) | Estimated time of preparation [h] | Chlorophorm Mabs (g g$^{-1}$) | Chlorophorm Vabs (cm$^3$ cm$^{-3}$) | Stress at 60% strain (kPa) | Reference |
|---|---|---|---|---|---|---|
| Swellable porous prepared with soft template method | 145 | ~0.6 | 32.3<br>21.5 (Dichloromethane) | 4.33<br>3.23 (Dichloromethane) | >300 | This work |
| Swellable porous PDMS | 144 | >12 | 34 | 4.14 | 5 | 24 |
| PDMS sponge | 120-130 | >5 | 11 | 1.34 | 12 | 23 |
| PDMS sponge | 138.9 | >8 | 12 (Dichloromethane) | 1.8 (Dichloromethane) | - | 26 |
| PDMS/MWNTs | 153.4 | ~40 | 20.5 | 4.24 | 330 | 13 |
| Bacterial Cellulose aereogel | 146.5 | >20 | 185 | 0.84 | ~20 | 37 |
| Melamine sponge/rGO | 132 | Different days | 165 | 1.25 | - | 38 |
| Nanofibrillated cellulose | 140 | >24 | 102 | 0.75 | - | 39 |
| Paper waste | - | 15 | 150 | 0.58 | - | 40 |
| rGo/PVDF | - | >18 | 20 | 0.028 | - | 41 |
| Graphene/EDA | 155 | 24 | 180 (DCM) | 0.84 (DCM) | 6 (50%) | 42 |
| MTMS–DMDMS gels | 152.6 | >24 | 14 | 2.54 | 10 | 43 |
| 3D graphene framework | - | >15 | 470 | 0.66 | 4 | 44 |
| Silanized melamine sponge | 151 | >1.5 | 163 | 1.94 | - | 45 |

On the other hand, PDMS porous material evidenced a lower mass-based absorption capacity with respect other porous oil absorbents due to the higher density of the materials. Consequently, PDMS porous material with higher mass absorption capacity and volume absorption capacity is on demand for practical application. Moreover, as evidenced before, high mechanical strength is on demand for practical application since the viable stack height of the sorbent after absorption is only a few of centimeters if this value is low. In table 1 the here prepared PDMS sponges are compared with other reported sorbents based on PDMS or other materials. Is evident as at 60% strain the material evidences one of the higher mechanical strength ever reported. As clearly visible the preparation time is considerably shorter with respect all the other sorbents. This is very important in case of emergency in which timely intervention is required. Due to the lower density of the material and large pores chloroform and dichloromethane mass absorption evidenced one of the higher value for PDMS based materials. Since mass-based absorption capacity is strongly affected by the density of absorbing materials, volume absorption capacity was also take in account to characterize the absorptive capability of the sponges, as is not a function of the solvent and absorbent density and better describe the absorption capacity of the material. Interestingly, to the best of our knowledge the best volume absorption capacity for oil absorption has been obtained in the present work making the system valuable for real applications. As described before, this could be due to increased porosity and the presence of larger pores with respect other PDMS based materials. However, larger pores can favor water penetration inside polymeric matrices. To evaluate the ability of hydrophobic PDMS to retain water we calculated the breakthrough pressure $P_B$ from $P_B = \rho g h_{max}$, where $\rho$ is density of water, g is gravitational acceleration and $h_{max}$ is maximum height of the water column that the PDMS sponge can support. Experimental results evidenced a $P_B$=1170Pa. Experimentally measured breakthrough pressures of porous PDMS is compared to theoretical $P_B$ ($P_{BT}$) of hydrophobic membrane with cylindrical pores using Laplace-Young equation:

$$P_{BT} = \frac{2\gamma |\cos(\theta_w)|}{d}$$

Where $\theta_w$ is water CA of the porous PDMS surface, $\gamma$= 0.0728 N m$^{-1}$ is the water surface tension of water air interface and d is the average pore diameters. The experimental $P_B$ is not in good agreement with theoretical value ($P_{BT}$=583 Pa). The difference can be attributed to the large variation in the pore dimension inside polymeric matrices and the presence of an elevated number of pores smaller than average pore dimension (<300μm) (figure S1)[46]. Nevertheless, the experimental $P_B$ is comparable with $P_{BT}$ that can be calculated for other PDMS porous material for water oil separation with pores more homogenously distributed for which values between 731 Pa and 1150 Pa can be calculated[25,26]. Moreover, highest water contact angle for sponges fabricated with only PDMS was observed, probably due to the absence of organic residual materials on the pores surface that can remain if hard template technique is used. This confirm as the here innovative synthetic procedure can produce materials with improved properties for water oil separation.

## Conclusions

In conclusion, a new route to fabricate 3D interconnected porous PDMS sponge has been proposed. The proposed fabrication technique is simpler, easier to be scaled up and dramatically faster than other reported porous absorbents. The obtained pores have a dimension range from tens of micrometres to millimeter. Moreover, different sponges prepared in different times evidenced replicable behaviour in dichloromethane absorbency suggesting as the synthetic route address to the fabrication of sponges with the same porosity. Owing to their hydrophobicity and superoleophilicity, the obtained PDMS sponge can selectively collect oils and organic solvents from water few seconds. The sponges are able to keep collected oil inside polymeric structure until a compression is applied and the porous PDMS is readily utilizable for a new cycle of oil collection. The here prepared sponges exhibit improved mechanical properties with respect to other PDMS absorbents. This permit to reuse the material hundreds of times without loss of function. The volume absorption capacity is higher than other porous materials reported for water oil separation and mass absorption capacity is comparable with the better values recorded on PDMS oil absorbents. Probably, as previously observed[23], the heterogeneity in pore dimension are crucial to obtain materials with high porosity and consequently higher sorption capacity and to prevent water penetration. The achieved volume and mass absorption capacity, the improved hydrophobicity and density are fully appropriate for practical applications allowing easy transport and storage.

As an example, for the absorption of 1 ton of petroleum (density ~ 0.88 g cm$^{-3}$), only ~ 56 kg of the materials are estimated to be required, that correspond to a sponge volume of only 0.28 m$^3$ (approximately equal to a volume occupied by ~ 3.6 persons). Improving the already good results and decreasing the costs with respect to more complicate systems previously reported based on PDMS and carbon nanotubes[13].

Moreover, we believe that the simple and dramatically faster preparation strategy can be successful in case of emergency permitting *in situ* fabrication and can offer inspiration for researchers to prepare other 3D interconnected porous materials.

## Acknowledgements

The support from Cohesion fund 2007-2013 - APQ Ricerca Regione Puglia "Programma regionale a sostegno della specializzazione intelligente e della sostenibilità sociale ed ambientale - FutureInResearch" under Grant no. 9EC1495 (Ultrasensitive sensor for food analysis) and Fondazione CARIPUGLIA, project: "Materiali innovativi porosi nanocompositi per la rimozione e il recupero di composti fenolici da acque di vegetazione olearie" are acknowledged. We thanks Dr. Marco Perrone for his technical help in absorption measursments.